\documentclass[article,twocolumn,amssymb,amsfonts,groupedaddress,floatfix]{revtex4}
\usepackage{amssymb,amsmath,graphicx,color}
\allowdisplaybreaks
\usepackage{amsfonts}
\usepackage{verbatim}

\newcommand{\bgt}{\begin{itemize}}
\newcommand{\ent}{\end{itemize}}

\newcommand{\lan}{\langle}
\newcommand{\ran}{\rangle}

\newcommand{\f}{\frac}

\newcommand{\bbm}{\begin{bmatrix}}
\newcommand{\ebm}{\end{bmatrix}}
\newcommand{\bes}{\begin{equation*}}
\newcommand{\ees}{\end{equation*}}
\newcommand{\be}{\begin{equation}}
\newcommand{\ee}{\end{equation}}
\newcommand{\beqy}{\begin{eqnarray}}
\newcommand{\eeqy}{\end{eqnarray}}
\newcommand{\beq}{\begin{eqnarray*}}
\newcommand{\eeq}{\end{eqnarray*}}

\newcommand{\bpm}{\begin{pmatrix}}
\newcommand{\epm}{\end{pmatrix}}

\long\def\symbolfootnote[#1]#2{\begingroup
\def\thefootnote{\fnsymbol{footnote}}\footnote[#1]{#2}\endgroup}

\begin{document}

\setkeys{Gin}{width=0.5\textwidth}

% This is for scaling the size of pictures to 100percent of a column width

\title{Direct spectrum analysis using  a threshold detector with application to a superconducting circuit}

\author{G. Ithier, G. Tancredi and P. J. Meeson}
%\author{G. Tancredi}
%\author{P.J. Meeson}

\affiliation{Department of Physics, Royal Holloway, University of
London, United Kingdom}

\begin{abstract}

We introduce a new and quantitative theoretical framework for noise spectral analysis using a threshold detector, which is then applied to
a superconducting device: the Cavity Bifurcation Amplifier (CBA).
We show that this new framework provides direct access to the environmental noise spectrum with a sensitivity approaching the standard quantum limit of weak continuous measurements. In addition,
 the accessible frequency range of the spectrum is, in principle, limited only by the ring down time of the CBA. This on-chip noise detector is non-dissipative and works with low probing powers, allowing it to be operated at low temperatures ($T<15$mK).
We exploit this technique for measuring the frequency fluctuations of the CBA and find a low frequency noise with an amplitude and spectrum compatible with a dielectric origin. 

%This method is applied to a Cavity Bifurcation Amplifier for measuring the noise spectrum
% of frequency fluctuations of the resonator: $S_{\delta f/f}$ when the circuit is biased at and the noise spectrum of
%magnetic flux. $S_{\delta f/f}$ has a $1/f$ scaling with an amplitude
% of $5.{10}^{-15}/$Hz at $1$Hz and with an average photon number
%of a few $10^2$ in the resonator. We give a theoretical estimate for the minimum noise floor of this technique
%and compare it to a linear measurement setup, assuming the same photon flow leaking out of the resonator.
%The noise floor is comparable to the photon shot noise limit and thus much smaller than the added equivalent thermal
%noise of a state of the art HEMT amplifier. 

\end{abstract}


\maketitle
%\begin{Scode}{echo=FALSE}
%\end{Scode}

\section{Introduction}
Due to their potential scalability, superconducting circuits provide a promising framework
for Quantum Information Processing. However, as solid state systems, they suffer from strong environmental
noise sources that limit their quantum coherence. Despite great improvement in coherence times over the recent years,
which is mainly due to clever optimized
designs~\cite{vion_manipulating_2002,koch_charge-insensitive_2007,schreier_suppressing_2008,
Paik3DCavityPRL2011}, 
and the proof of principle of a correction algorithm on a quantum memory~\cite{reed_realization_2012}, 
the quantum coherence times are not yet sufficient for the realization of non-trivial fault-tolerant quantum computations~\cite{NielsenChuang}.
As environmental noise sources cause decoherence their extensive characterization is a key issue in order
 to identify the origin of the noisy subsystems and improve
materials properties, minimize coupling to these sources with better designs, or implement
 special dynamical decoupling sequences~\cite{bylander_noise_2011}.
% It would require $4$ characteristics to be achieved: high sensitivity,
%low temperature of operation, low probing power, and most importantly  the
%frequency dependence of environmental noise sources
Until recently, all characterization techniques (except notably \cite{PosterCMMP11,Yan2012}) of noise sources in superconducting quantum bits circuits used the decay of coherence functions borrowed from NMR~\cite{yoshihara_correlated_2010,ithier_decoherence_2005,bylander_noise_2011}.
%: either Ramsey, Hahn echo decays, multipulse control
%sequences~\cite{collin_nmr-like_2004,bylander_noise_2011}, in the free evolution regime or Rabi decays
% in the driven regime.
% averaged over several realizations of the driving pulse sequence.
These techniques are the most sensitive and they operate at low temperature ($T<50$mK) and low probing power,
however they do not give direct access to the frequency dependence of the noise.
 % but to its convolution with some kernel
%function which acts as a frequency domain filter and whose properties depend on the pulse sequence.
Indeed, from the dependence of the decay functions with the control parameters (like charge or magnetic flux),
they give access to the standard deviation of the noisy control parameter integrated
over a frequency window which is, for example, the
bandwidth of the acquisition process in the case of a Ramsey sequence~\cite{ithier_decoherence_2005,yoshihara_correlated_2010}.
%Other standard noise measurement techniques do not achieve the $4$ requirements altogether
%(high sensitivity, low T, low driving power, and frequency dependence): for instance SQUID
% amplifiers for flux noise measurements are limited at temperature higher than $50$mK because of dissipation in
%the shunt resistors, Single Electron Transistors for charge noise measurement have a low bandwidth of operation,
% Kinetic Inductance Detectors are interesting probes of dielectric noise but they need a high driving power.
% In ref\cite{Yan2012}, authors presentfor accessing the environment noise spectrum of a flux quit

In this article, we first discuss the standard operation of a CBA, to motivate and introduce the theoretical framework required for using a 
threshold detector as a spectrum analyzer.
%~\cite{Yan2012}.
 Then we apply this technique to the measurement of the
 frequency fluctuations of a CBA.
 We demonstrate that this method combines the advantages of state of the art noise measurement techniques in superconducting circuits \cite{gao_noise_2007,Lindstrom2011,SQUIDHandbook} with the advantages of non dissipative quantum bit readout setups, achieving the four following aims together:
a high bandwidth given by half the repetition rate of the measurement,
a high sensitivity which may approach the standard quantum limit of a weak continuous measurement,
a low temperature of operation ($<15$mK) due to the absence of on-chip dissipation and a low probing energy (in our case $\approx 10^3$ photons in the CBA). This technique can be applied to any detector involving a threshold effect and in particular qubit state measurement setups.

%present a novel technique for accessing environmental noise spectra which can be applied to
%any threshold detector and in particular quantum bit state measurement.
%We provide the theoretical framework 
%We first introduce a general theoretical framework for implementing spectral analysis with threshold detectors and

%We note that the measurement method is analogous to an analog to digital conversion
%with $1$ bit of resolution and in presence of an additive dither.

\section{The Cavity Bifurcation Amplifier}

\begin{figure}
  \includegraphics[width=8cm]{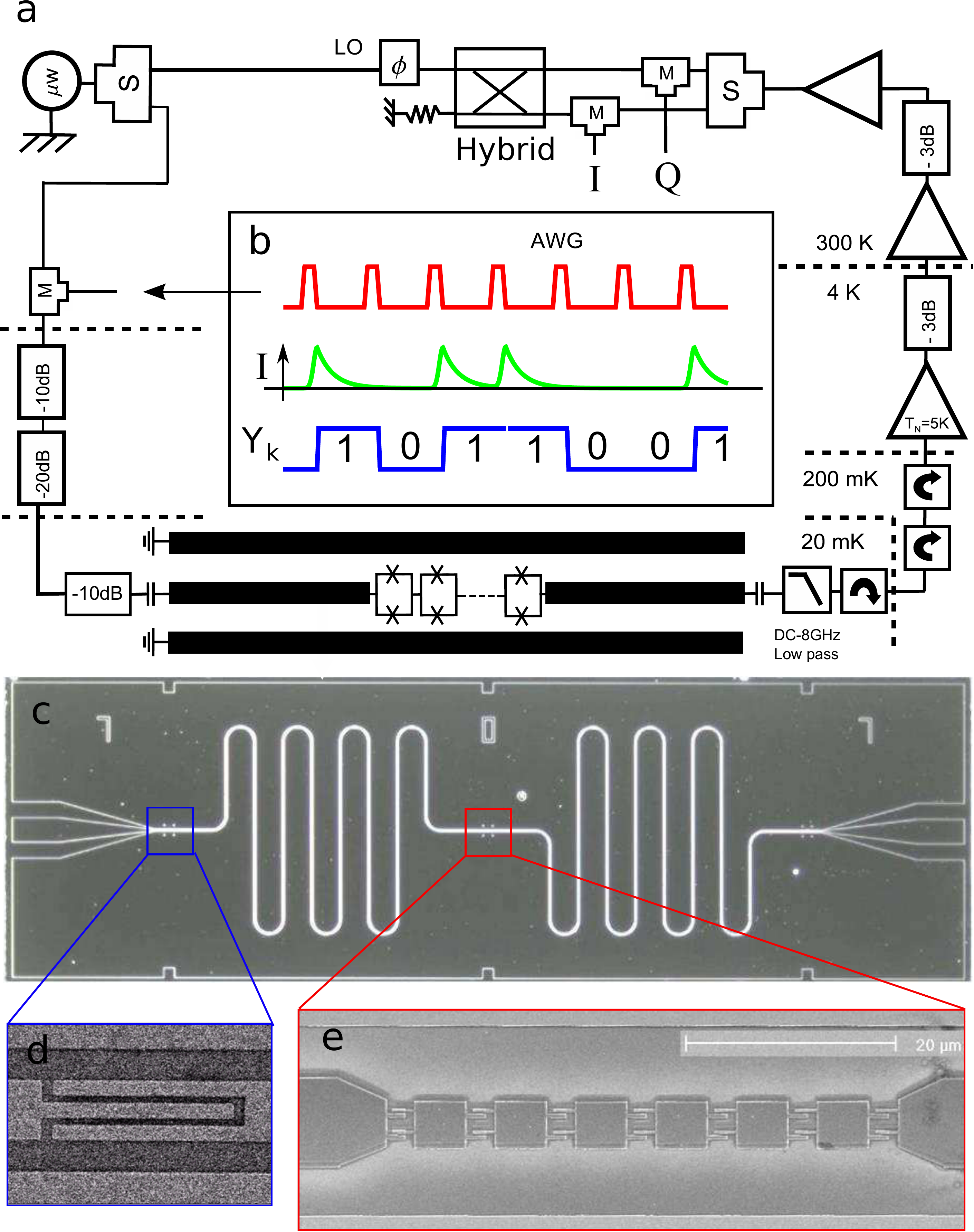}
  \\
  \caption{  \textbf{Experimental Setup}. \textbf{a)} Homodyne detection setup. \textbf{b)} Waveforms: trapezoidal pulses
  shaping the microwave applied to the sample (red). One quadrature of the demodulated signal shows switching events as
  random jumps (green). This switching signal is recorded as a binary signal (blue).
   \textbf{c) } Sample chip, made of a $200$nm thick niobium coplanar waveguide deposited on an oxidised silicon wafer.
   \textbf{d)} Interdigitated input capacitor. \textbf{e)} Array of $7$ aluminum SQUIDs located at an anti-node
   of the electric current distribution of odd harmonics.  }
  \label{Fig1}
\end{figure}

The CBA detector has been extensively studied for the purpose of superconducting quantum
bit readout~\cite{mallet_single-shot_2009, vijay_invited_2009,boulant_quantum_2007,siddiqi_dispersive_2006}.
Our device consists of a section of superconducting niobium coplanar waveguide
%of length $l=30.0$mm
enclosed between input and output capacitors which provide coupling to this resonant Fabry-Perot like structure
 (see Fig.\ref{Fig1}, more details are given in \cite{Tancredi1,palacios-laloy_tunable_2008}).
An array of Superconducting QUantum Interference Devices (SQUIDs) is located in the middle of this structure at anti-nodes
of the electric current distribution of odd numbered harmonic modes. This setup provides a strong and tunable non linearity for these modes.
 Due to this non linearity, this system exhibits parametric amplification
below a critical number $n_c$ of photons populating the resonator
 and a bifurcation phenomenon above it (here $n_c\approx 250$ for the third harmonic mode 
we consider in the following).
This bifurcation is a transition between two dynamical states of oscillation: one
of small and one of large amplitude, which can be easily detected using commercially available 
cryogenic
amplifiers. The transition rate between these two states depends on experimental parameters (written generically as variable $X$) which are for instance the resonant frequency of
 the mode, its quality factor, the frequency and amplitude of the microwave driving. Some of these experimental parameters may be controlled and some others might be the subject of random fluctuations,
 which can be detected by the CBA. 

As a first characterization of the sensitivity of this system as a detector, we repeatedly probe the CBA with microwave driving pulses
(duration $\tau = 35\mu$s $\approx 10 /\gamma$ where $\gamma$ is the linewidth of the mode, in order to damp any transient) at a given rate $\nu_{rep}$
(typically $5$kHz), and we record the state of the resonator at the end of the  driving pulse (labeled $Y_k$ at time step 
$t_k=k/\nu_{rep}$) in binary format ($Y_k=1$ for the high amplitude
state, $Y_k=0$ for the low amplitude state). Counting over $N=10^3$ events, 
one obtains the average switching probability $p_{exp}=1/N\sum_{k=1}^N Y_k$ for a given set of parameters $X$. Actually, as discussed in the following, $X$ might undergo random fluctuations, meaning that
the experimentalist can control only the average value of $X$: $\lan X\ran$.
 Recording the switching probability while ramping the control parameter  $\lan X \ran$ across the bifurcation frontier provides the switching probability curve or "S-curve" whose $10\%-90\%$ width $\Delta X$ defines its sensitivity to fluctuations of $X$, that is, a shift of $\Delta X$ can be detected within a single probing pulse with a high level of confidence.
 
The natural question which arises now is: Is it possible to infer more information from the array $\{Y_1,Y_2,...,Y_N\}$ than just the average switching probability $p_{exp}$? 
%and detect for instance fluctuations of $X$ smaller than $\Delta X$ on short time scales (of order $1/\nu_{rep}$).
 % This slope gives the frequency to probability conversion: $\delta p = \partial p /\partial \nu \delta \nu$
A first improvement is to measure the average switching probability $p_{exp}^n$ over
 subsets of $n$ events. This allows the detection of fluctuations of $X$ of order $\Delta X/\sqrt{n}$ still with a high level of 
confidence but with a much lower bandwidth of the order of $\nu_{rep}/n$. 
 Experimentally, we do observe fluctuations of the switching probability $p_{exp}=p_{exp}^N$  which are 
 well above the expected statistical noise ($\approx 1/\sqrt{N}$ where $N=10^3$ events), indicating
a low frequency noise present in the experimental parameters. In addition, the measured experimental 
value for the switching curve as a function of the frequency of the mode: 
$\Delta \nu \approx 4.5$kHz is greater by a factor $2$ from the theoretical prediction
obtained from the Dykman model~\cite{QuantumBifurcationDykman1988}.
%\begin{equation}\label{SwidthPrediction}
%\frac{\Delta \nu}{\nu_0}=\frac{3^{1/3}}{2^{4/3}}\frac{1}{Q N_c^{2/3}}
%\end{equation}
%where $Q$ is the quality factor of the resonator, $N_c$ the critical number of photons populating the resonator for the bifurcation to start.
 Both facts indicate that a non-negligible part of the switching curve width is due to fluctuations of the experimental
parameters. We are thus led to consider a 'doubly stochastic process': the outcome of the
detector is a random process $Y_k$ depending on a switching probability which is \textit{itself} a random
process (since it depends on a noisy parameter $X$). After checking the noise level of our microwave setup,
we can exclude frequency fluctuations of the probing pulses and microwave amplitude
fluctuations at the level of the sample. The most likely origin of the noise source is microscopic and on-chip. Such fluctuations may be characterized as inducing fluctuations in the resonant frequency of the cavity.
% which account for a maximum fraction of $\approx 1/10$ of the width of the Scurve.

We will now show that it is possible to extract the spectrum of the frequency fluctuations of the cavity from
the binary array $\{Y_1,Y_2,...,Y_N\}$.
For this purpose, we need to introduce a general model for our detector (which, we note, can be applied to any system involving a threshold effect).

\section{Modeling of a threshold detector}
\begin{figure}
\includegraphics[width=8cm]{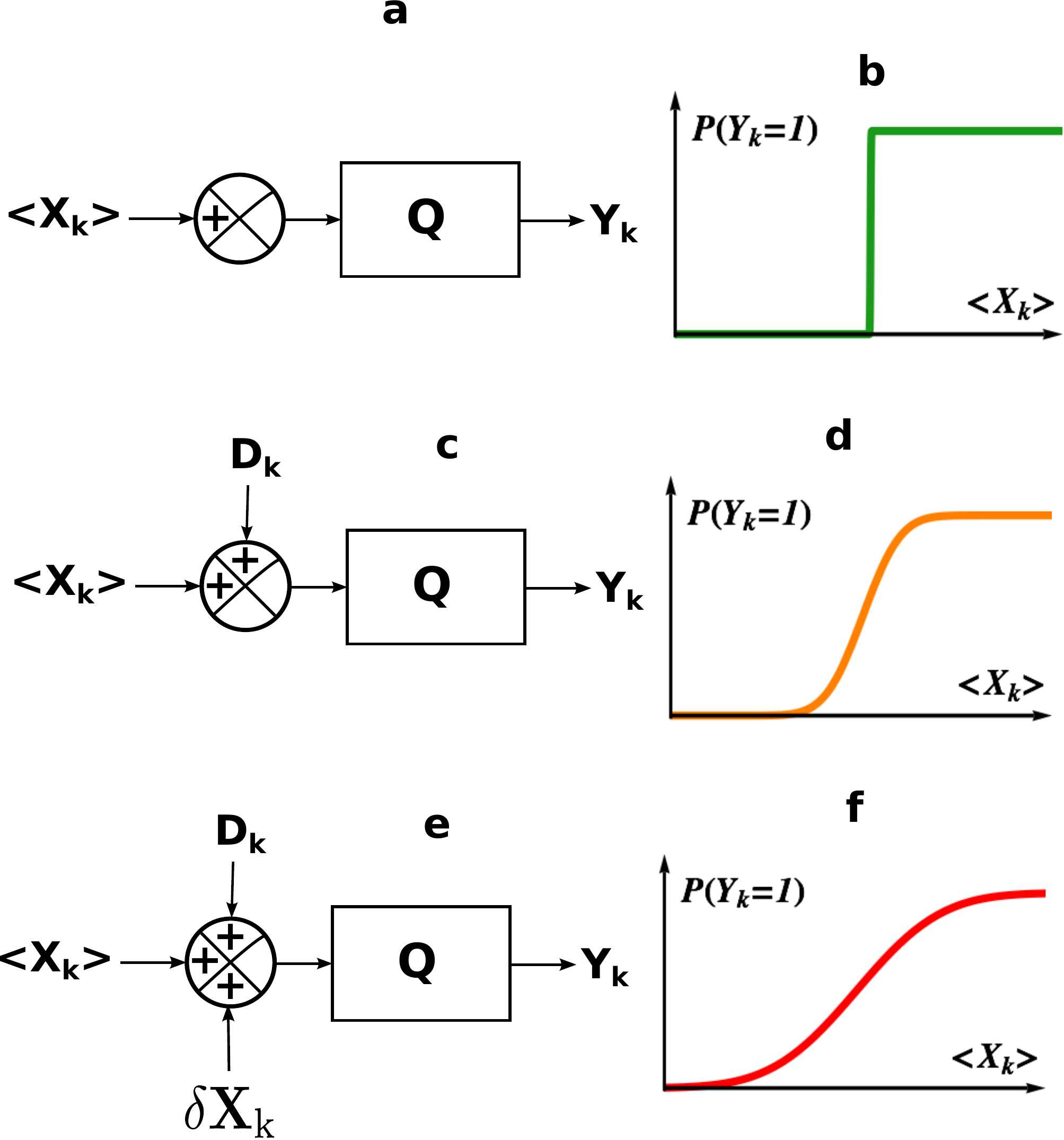}
\caption{ \textbf{Modeling of a threshold detector as a $1-$bit analog to digital converter}.
\textbf{a)} Ideal threshold detector: in the absence of any noise source, the outcome $Y_k$ at time $t_k$ is modeled as the result of the $1-$bit analog to digital conversion of the input signal sampled at time $t_k$: $X_k$ considered constant 
and equal to $\lan X \ran $: $Y=Q[X-x_0]$. The quantizer threshold is set
to $x_0$. 
\textbf{b)} Response function of the detector: $P(Y_k=1)$ plotted as a function of $\langle X \rangle$ shows a "sharp" threshold.
\textbf{c)} In order to model the finite sensitivity of the detector, a random noise $D$ (here a set of independent gaussian distributed random variables $D_k$) is added to the input signal $X_k$ prior to thresholding. 
 \textbf{d)} The response function is now broadened by the $D_k$ fluctuations.
\textbf{e)}: The input signal $X_k$ undergoes itself some random fluctuations which are modeled by the
addition of the $\delta X_k$ random variable to $\lan X\ran$.
\textbf{f)}: The response function is broadened by both fluctuations of $D_k$ and $\delta X_k$.  }
 \label{Fig2}
\end{figure}

We consider the CBA 
%(or any threshold detector)
 as a generic bistable system:
its state labelled $Y$ is a random variable which can take two different values ($Y=1$ or $0$) with probabilities dependent on whether some parameter
  $X$ is above or below a threshold value $x_0$. By offsetting $X$, we set $x_0$ to $0$ in the following. 
Considering first the ideal case where thermal and quantum noises are \textit{absent} from this detector, we
have a "sharp" threshold: $Y=0$ with certainty if $X<0$ and $Y=1$ with certainty if $X>0$. In this case, this detector is completely analogous to a $1-$bit analog to digital converter (see Fig.\ref{Fig2}a):
$Y=Q[X]$ where $Q$ is a digitizer function ($Q[X]=1$ if $X>0$ and $Q[X]=0$ if $X<0$). Now,
consider a time varying $X_t$ named an 'input' signal, which is sampled by this threshold
detector at regular time intervals $t_k=k/\nu_{rep}$ ($k\in \{1,..,N\}$ to obtain a binary array of outcomes $\{Y_1,Y_2,...,Y_N\}$.
What can be inferred about the temporal variations of $X_t$ from the $\{Y_1,Y_2,...,Y_N\}$ array?
Provided that the signal has some frequency components only for $f<\nu_{rep}/2$ (the Shannon criterion) and that the mean value 
$\langle X\rangle$ of the signal is within a fraction of its standard deviation $\sigma_X$ of the threshold, then the answer is: only some crude information about the largest Fourier component of $X_t$ can be inferred.
For instance, a sinusoidal input oscillation of $X$ around the threshold will be converted into a rectangular
output, thus corrupting the spectrum of $Y_k$ with harmonic generation (see Fig.\ref{TimeDomainFFT}a and
  Fig.\ref{TimeDomainFFT}b).
%However, we will now see that using a technique similar to noise engineering, it is in fact possible recover significant amounts of information about the input signal $X_t$ from the binary array $\{Y_1,Y_2,...,Y_N\}$. 
%Being a strongly non linear process, the digitization of the input signal introduces a distortion which contains harmonics of the $X_t$ frequency components.

However, in analog to digital conversion systems, it is common practice 
to add a small amount of noise to the input signal $X_t$ prior to digitization (see Fig.\ref{Fig2}c) in order to control or "shape" the associated distortion. For instance, this technique is implemented in fast oscilloscopes where $X_t$ is an input voltage and the noise is generated by the pre-amplifier stage of the scope. A careful engineering of this input noise can increase the effective resolution of the converter at the expense of the sampling rate: this is the so-called "oversampling" technique\cite{hauser1991principles}.
% in the presence of an additive random noise $D_k$
% threshold broadened by the additive noise $D_k$ (orange), threshold broadened by $D_k$ and $X_k$ fluctuations (red). 

We will now consider an analogous situation for our threshold detector and focus on its implication for the spectrum of the digitized signal. 
A random variable $D_k$ is added to the input signal prior to thresholding (see Fig.\ref{Fig2}c), such that the output signal is now $Y_k=Q[X_k+D_k]$, where $X_k=X_{t_k}$ is the input sampled at time $t_k$.
We assume that the $D_k$ variables are independent and generated from a stationary random process with zero mean and probability density $P_D(d)$. We focus first on the statistics of order one of this probabilistic model.
% and see how the
% extra random variable $D_k$ is going to broaden the detection threshold of this quantizer.

% then it is possible to show the following results\cite{wannamaker_theory_2003,lipshitz_non_substractive_dither}:
%\begin{itemize}
 %   \item 
 
%%%%%%%%%%%%%%%%%%%%%%%%%%%%
 \subsection{First order statistics of the model}             %
%%%%%%%%%%%%%%%%%%%%%%%%%%%%

 For a \textit{single sampling} of the input signal $X_k$ at a given time $t_k$ having the value $x$,  the outcome of the digitization process $Y_k$ takes the value $1$ with probability $p(x)$ and the value $0$
with probability $1-p(x)$ where $p(x)$ is 
the conditional probability:
\be
p(x) = \mathbb{P}\left( Y_k=1/X_k=x  \right) 
\ee
to observe $Y_k=1$ knowing that $X_k=x$.
This probability can be related to the cumulative distribution of the $D_k$ variables, $P_D(d)$:

\begin{eqnarray*}\label{Proba}
p(x) &=& \mathbb{P}(X_k+D_k>0/X_k=x) \\
&= &\mathbb{P}(D_k>-x)=\int_{-x}^{\infty} P_D(u)du 
%\\ &=&\left( P_D \otimes H \right) (x)
\end{eqnarray*}

% and $H(u)$ is the Heaviside function  ($H(x)=0$ if $x<0$ and $H(x)=1$ if $x>0$ which is the response of the ideal quantizer).
%Here the event we consider is $\{Y=1\}/\{X=x\}$ which is equivalent to $\{X+D>0\}/\{X=x\}$ or $\{D>-x\}$.
As a result of the $D$ fluctuations, the "sharp" threshold of the ideal quantizer
is broadened by the distribution $P_D$ (see Fig.\ref{Fig2}d).
However, experimentally $X_t$ can fluctuate over time so we cannot access directly $p(x)$. Instead, one is measuring an average probability $p_{exp}$ calculated over many probing pulses (at $t_1,t_2,...,t_N$): 
\begin{equation}
p_{exp}  =  \frac{1}{N} \sum_{k=1}^N Y_k 
\end{equation}
which, in the limit of the law of large numbers, can be approximated by:
\be
\approx \frac{1}{N} \sum_{k=1}^N  \langle Y_k \rangle 
\ee
where $\langle Y_k \rangle = \mathbb{P}(Y_k=1)$.
So what is the value of $p_{exp}$ knowing that $X_t$ can fluctuate over time?    
To answer this question, we need to assume two more hypotheses on the process $X_t$:
first,  $X_t$ undergoes a random stationary process with a distribution probability $P_{X}$ centered around an average value 
$\lan X \ran$ which can be controlled experimentally (like an average magnetic flux or an average gate voltage). Then we need to assume a quasi-static approximation: the fluctuations of $X_t$ should be slower than the sampling time, (i.e. the duration of a single microwave probing pulse in the case of the CBA). 
With these two hypotheses, $\mathbb{P}(Y_k=1)$ does not depend on $k$ and is the average of $p(x)$ 
weighted by the distribution of $X$:

\begin{eqnarray*}
p_{exp} & \approx & \mathbb{P}(Y_k=1) =  \int_{-\infty}^{+\infty} \mathbb{P}(Y_k=1/X_k=x) P_X(x)dx \\
&= &\int_{-\infty}^{+\infty}p(x)  P_X(x)  dx .
\end{eqnarray*}
Setting  $X_t=\delta X_t + \langle X \ran$, we can rewrite $p_{exp}$ as a function of $\lan X \ran$:

\be
p_{exp} (\langle X \ran) \approx 
 \int_{-\infty}^{+\infty}  p(x) P_{\delta X} (x-\langle X \ran) dx =
 p \otimes P_{\delta X} (\langle X \ran) 
\ee

 The experimental probability $p_{exp}$ of detection considered as a function of the control parameter 
 $\langle X \ran$ is thus the convolution of the response of the detector $p(x)$ with the probability distribution of $X$. 
 The $p(x)$ response curve of the detector, already broadened by the fluctuations of the $D_k$, is further broadened by the fluctuations of the $X_t$ process (see Fig.\ref{Fig2}f). 
\begin{comment}
\begin{eqnarray}
 & =& \mathbb{P}(x+d>0)=\mathbb{P}(\bar{x}+\delta x +d>0)  \\
    &=& \mathbb{P}(\delta x +d> - \bar{x})   \\
    &=&    \int_{-\bar{x}}^{\infty} du \int du' P_X(u')P_D(u-u') \\
    &=&   \int du' P_X(u') \int_{-\bar{x}}^{\infty} du  P_D(u-u') \\
    &=&   \int du' P_X(u') \int_{-u'-\bar{x}}^{\infty} dy  P_D(y) \\
    &=& \int du' P_X(u') p(u'+\bar{x}) \\
    & =& (p\otimes P_X)(\overline{x})
\end{eqnarray}
\end{comment}
The threshold of the digitization process is no longer "sharp", it has an '$S$ like' shape with
%some finite width coming from both the fluctuations of $D_k$ and $X_k$ from sample to sample (see Fig.(\ref{Fig2}d)). 
a $10\% -90\%$ width (defined as $\Delta X$), which can be related to the standard deviations of $X$ and $D$:
 $\Delta X \approx 2.56 \sqrt{\sigma_D^2+\sigma_X^2}$ in the case of gaussian distributions for 
 $D_k$ and $X_k$. We will see that this relation is useful for calibrating our detector. 
 
 Having studied the first order statistics of our detection model, we now focus on the second order statistics and demonstrate that the spectrum of the $X$
parameter can be extracted from the experimental binary array $\{Y_1,Y_2,...,Y_N \}$.

%%%%%%%%%%%%%%   Commented section
\begin{comment}
Eq.\ref{Proba} can be linearized in the case where $\sigma_X < \sigma_D$. One gets: 
\begin{eqnarray}
\delta p(x) & = & p(x)-\frac{1}{2}=\int_{-x}^{\infty} P_D(u)du-\frac{1}{2} \\
  & \approx & \delta x . P_D(0)= 
\frac{\delta x}{\sqrt{2\pi} \sigma_D}
\end{eqnarray}
where $\delta p=p-1/2$, $\delta x=x-\overline{x}$ and for
$D_k$ variables with a gaussian distribution:
\begin{equation}
P_D(d)=\frac{1}{\sqrt{2\pi} \sigma_D}e^{-\frac{d^2}{2\sigma_D^2}}
\end{equation}
$\delta p$ can be considered as the input signal $\delta x$ normalized by
 the standard deviation of the noisy $D$, and it will be useful in the following to use it as an auxiliary input.
 \end{comment}
 %%%%%%%%%%%%%%%   End commented section

 %%%%%%%%%%%%%%%%%%%%%%%%%%%%%%%%%%%%%%%%%%%%%%%
 \subsection{Second order statistics of the model: autocorrelation and spectral density}          %
 %%%%%%%%%%%%%%%%%%%%%%%%%%%%%%%%%%%%%%%%%%%%%%%
 
We consider the autocorrelation of the $\{Y_1,Y_2,...,Y_N\}$ array and show here that it can be related to the autocorrelation of the $X_t$ process.
 We define first the fluctuation $\delta Y_k =Y_k-p_{exp}$. 
 We know that  $\langle Y_k \ran=p_{exp}$ and $\langle Y_k^2\ran=\langle Y_k\ran$ (since 
 $Y_k$ takes only two values $0$ or $1$). As a consequence the variance $\sigma_Y^2$ of the $Y_k$ process is:

 \be
 \sigma_Y^2 = \langle \delta Y_k^2\ran  =p_{exp}(1-p_{exp})
 \ee
 
 %where $p_{exp}$ is the average probability: $p_{exp}=1/N\sum_{k=0}^N Y_k $. 
Then for $q \neq 0$, we have:
 
 \begin{eqnarray}
 \langle Y_{k+q} Y_k\rangle &=& \mathbb{P}(\{Y_{k+q}=1\} \, \textrm{AND} \, \{Y_k=1\}) \nonumber \\
   &=&\mathbb{P}(\{X_{k+q}+D_{k+q}>0\} \, \textrm{AND} \, \{X_k+D_k>0\})    \nonumber \\
   &=& \int_{0}^{\infty} \int_0^{\infty} P_{U,U} (u_1,u_2) du_1 du_2
   \label{AutoCorr3}
 \end{eqnarray}
 
 where $\mathbb{P}(\{Y_k=1\} \, \textrm{AND} \, \{Y_{k+q}=1\})$ is the joint probability to have the events $Y_k=1$ and
 $Y_{k+q}=1$. $P_{U,U}$ is the joint probability density of $U_1=X_k+D_k$ and
 $U_2=X_{k+q}+D_{k+q}$. Such a probability density is the double convolution of the joint probability of $X$: $P_{X,X}$, with the joint probability of D: $P_{D,D}(d_1,d_2)$ :
 
 \be
 \label{DoubleConvol}
 P_{U,U}(u_1,u_2) =
 \left( P_{D,D} \otimes P_{X,X} \right) (u_1,u_2)
% \iint_{\mathbb{R}^2}  dv_1 dv_2 P_D(v_1) P_D(v_2)
 %  P_{X,X}(u_1-v_1,u_2-v_2)  
 \ee
  Because the two random variables $D_1$ and $D_2$ are assumed to be independent and identically distributed, we have that $P_{D,D}(d_1,d_2)=P_D(d_1).P_D(d_2)$.
 \begin{comment}
 which gives the autocorrelation of $Y_k$:
 \begin{widetext}
\[
 \langlegle  Y_k.  Y_{k+q}\rangle = \int_0^{\infty}du_1 \int_0^{\infty}  du_2   \iint_{\mathbb{R}^2} dv_1 dv_2 P_D(v_1) P_D(v_2)
   P_{X,X}(u_1-v_1,u_2-v_2)  
\] \label{Dconvol}
\end{widetext}
\end{comment}
To go further we need to make more assumptions about the statistics of the $X_t$ process.

A gaussian hypothesis for $X_t$ is physically reasonable since we are dealing with a condensed matter system where sources of noises involve \textit{a priori} large numbers of uncorrelated fluctuating subsystems. In addition, since we are interested only in the second order statistics, such a gaussian hypothesis for $X_t$
is all that is necessary. Finally, this hypothesis will provide analytical formulas for the relation between
% $\langle \delta Y_{k+q} \delta Y_k \ran=\langle Y_k.Y_{k+q}\rangle-\langle Y\rangle^2$ and $\lan \delta X_{t_{k+q}} \delta X_{t_k}
% \ran $.
the autocorrelations of $Y_k$ and $X_k$.
We thus assume that the $X_t$ process is stationary and has the following joint probability density:
 
 \begin{equation}
 \label{XDensity2}
 P_{X,X}(x_1,x_2;t)=\frac{1}{2\pi \sigma_X^2 \sqrt{1-\rho^2_X}} 
e^{-\dfrac{(x_1^2+x_2^2 -2\rho_X x_1 x_2)}{2(1-\rho_X^2)\sigma_X^2}}
\end{equation}
where $\rho_X(t)=\langle  \delta X_t \delta X_0 \rangle /\sigma_X^2$ is the normalized autocorrelation 
(in Eq.\ref{XDensity2}, the dependence of $\rho_X$ on time is omitted for brevity).
 
\begin{figure}
  \includegraphics[width=8cm]{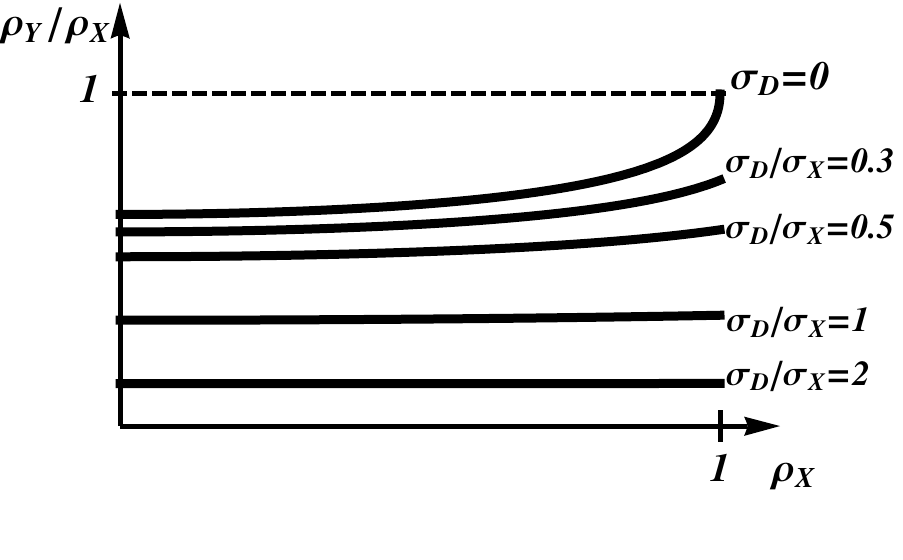}
  \\
  \caption{ The transfer function of the autocorrelation: $\rho_Y/\rho_X$ as a function 
  of $\rho_X$. For $\sigma_D=0$, there is no additive noise prior to thresholding, as a result, the transfer function is highly non linear. But as $\sigma_D/\sigma_X$ increases, the non linearity decreases, and for $\sigma_D > \sigma_X$ the transfer function is quasi-linear.}
 \label{TransfertFunction}
\end{figure}

The double convolution of Eq.(\ref{DoubleConvol}) and the integration in Eq.(\ref{AutoCorr3}) can then be calculated analytically and gives the main result of this paper: a direct relationship between the normalized autocorrelations of 
$X_k$ and $Y_k$: $\rho_X$ and $\rho_Y$,

\begin{eqnarray}
\rho_X(t_q)=\f{\langle  \delta X_{k+q} \delta X_k \rangle}{ \sigma_X^2 }, \qquad \rho_Y(t_q)=\f{\lan \delta Y_{k+q} \delta Y_k \ran}{\sigma_Y^2  } \\
\rho_Y(t_q) = \frac{2}{ \pi} 
\arctan{\left(\frac{\rho_X(t_q) }{ \sqrt{(1+\f{\sigma_D^2}{\sigma_X^2})^2 -\rho_X(t_q)^2 }} \right)}
\label{AutoCorrXY}
\end{eqnarray}
%with $\sigma=\sigma_D/\sigma_X$ and 
%for $t_q \neq 0$ 
Note that Eq.\ref{AutoCorrXY} is valid for $q >0$ only: two \textit{different} samples have to be considered
(when $q=0$, there is no relation between $\lan \delta Y_k^2 \ran=p_{exp}(1-p_{exp})$ and $\lan \delta X^2 \ran$). 
 This point will be important when considering the Fourier transform of the autocorrelation to obtain the spectrum.
It is then useful to define a transfer function $\rho_Y/\rho_X$ which is plotted as a function of $\rho_X$ in
 Fig.(\ref{TransfertFunction}) for different values of $\sigma_D/\sigma_X$. We consider the two limiting cases: 
 \begin{itemize}
 \item When $\sigma_D \to 0$, the additive noise $D$ disappears, and Eq.\ref{AutoCorrXY} simplifies to:
 \be
\rho_Y(t_q)=\f{ \lan \delta Y_{k+q} \delta Y_k  \ran}{\sigma_Y^2}= \frac{2}{\pi} \arcsin{(\rho_X(t_q))}.
 \ee
 This is a strong non linear relationship between $\lan \delta Y_{k+q}  \delta Y_k \ran$
 and $\lan  \delta X_{k+q} \delta X_k  \ran$, which accounts for harmonic generation.
 
 \item The other limit case is the more interesting one: when $\sigma_X \lesssim \sigma_D$,  one finds
 a quasi-linear relation between the autocorrelation of $X$ and the autocorrelation of $Y$:

 \be
 \label{XYrelation}
 \lan\delta Y_{k+q}  \delta Y_k \ran \approx 
  \f{2}{\pi} \left(\f{ \sigma_Y}{\sigma_D}\right)^2 \lan \delta X_{k+q}  \delta X_k \ran
 \ee
 valid for $q>0$.  
  \end{itemize}

%This relation allows to infer directly the autocorrelation of the input signal $X$ from the autocorrelation of the binary output array $\{Y_1,Y_2,...,Y_N\}$, which is a priori not obvious at all.
The linearity of Eq.\ref{XYrelation} gives a direct access to the autocorrelation of $X_k$ from the experimentally measured autocorrelation of the $\{ Y_1,Y_2,...,Y_N\}$ array. The harmonic distortion due to the thresholding is suppressed by the addition of the noise $D$ to the input signal. This is done at the expense of the 'gain' $\rho_Y/\rho_X$ which decreases as $\sigma_D$ increases. There is thus a tradeoff between linearity and gain.

 We will assume in the following that our detector operates in the regime $\sigma_X \lesssim \sigma_D$ which provides a good compromise between linearity and gain (for instance, when $\sigma_D = \sigma_X$, the gain is $\rho_Y/\rho_X \approx 0.32$ and is constant within $\pm 1\%$).
The gain $\frac{2\sigma_Y^2}{ \pi \sigma_D^2}$ can be easily calibrated: when the repetition rate increases, the correlation between two successive outcomes $\lan \delta Y_{k+1} \delta Y_k \ran$ converges to:
%$ \langle\elta Y_k \delta Y_{k+1}\rangle$ does \textit{not} converge
%to $\langle\delta Y_k^2\rangle=p_{exp}(1-p_{exp})$
\be
\label{Calib}
\lim_{\nu_{rep} \to \infty} \lan \delta Y_{k+1} \delta Y_k \ran =  \frac{2}{\pi} \f{ \sigma_Y^2}{\sigma_D^2} \sigma_X^2
\ee
Combined with the measurement of the switching curve width $\Delta X \approx 2.56\sqrt{\sigma_D^2+\sigma_X^2}$, Eq.\ref{Calib} provides estimates of the values of $\sigma_D$ and $\sigma_X$.
%The autocorrelation of $Y_k$ thus make a sharp separation between the correlations
%due to the digitization noise of the discrete outcome of the measurement (which appear only for $k=q$) and the underlying
%random process causing $\lambda$ (and thus $p$) to fluctuate.

\begin{figure}
\includegraphics[width=9cm]{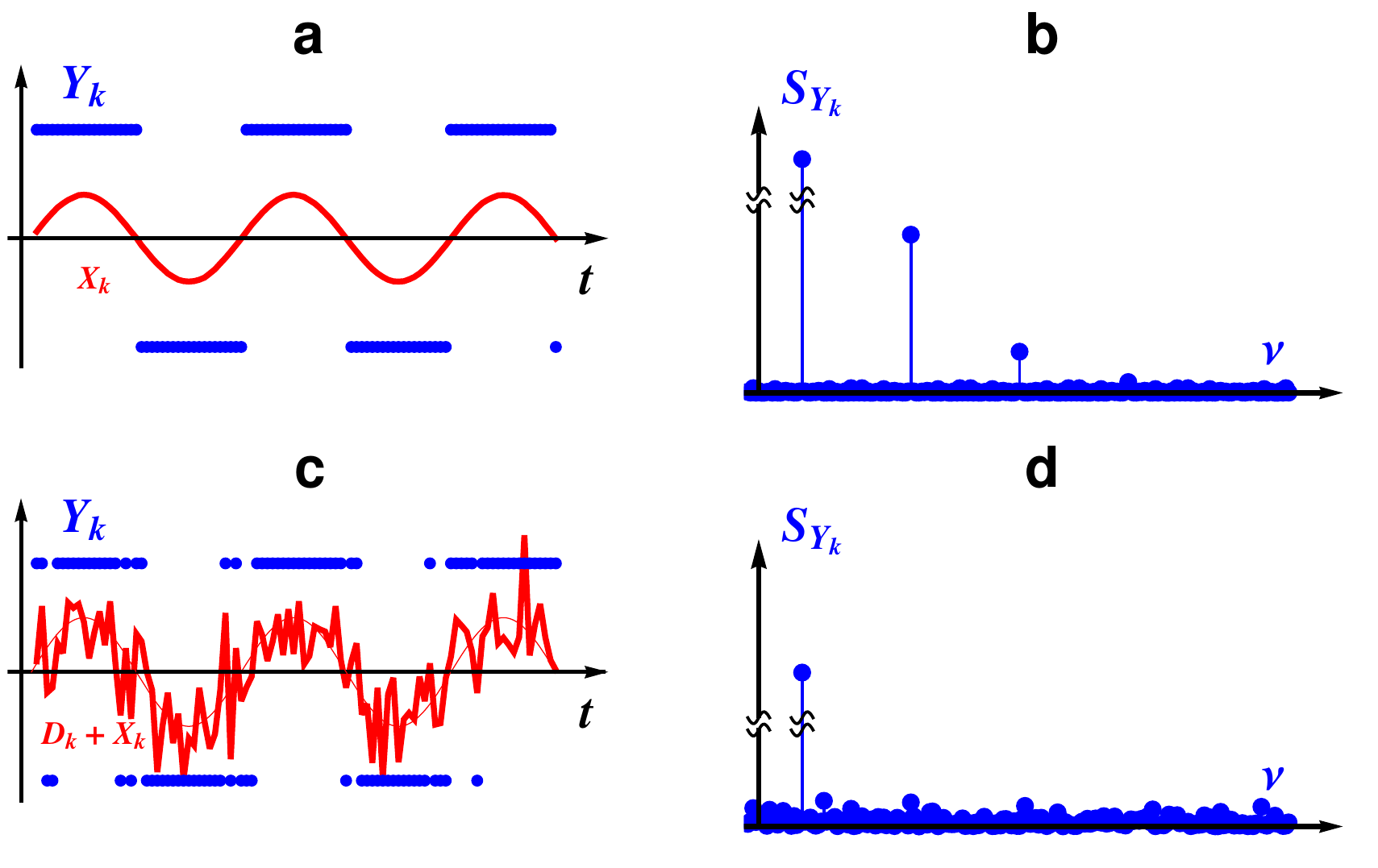}
\\
\caption{ \textbf{A numerical model showing the effect of an additive noise on the digitization process: example of a sinusoidal input}.
 \textbf{a)} Digitization of a sinusoidal input with a sharp threshold detector,  red: pure sinusoidal modulation around a threshold,
 blue: digitized samples (here $N=100$).
 \textbf{b)} Power spectrum of the digitized signal: the distortion created by the sharp thresholding appears
  as harmonics of the sinusoidal input.
\textbf{c)}: A random noise $D_k$ is added to the input signal $X_k$ prior to thresholding, such that $\sigma_D =\sigma_X$. Note here that the sinusoidal input is obviously not a gaussian distributed process. 
\textbf{d)}: Spectrum of $Y_k$: the higher harmonic content due to digitization has been "shaped" as a white noise background. 
Vertical and horizontal scales are the same for plots (b) and (d). }
 \label{TimeDomainFFT}
\end{figure}

Eq.\ref{XYrelation} can be rewritten in the frequency domain: by Fourier transforming Eq.(\ref{XYrelation}) and using the Wiener-Khinchin theorem, 
one obtains  the relation between the spectral densities of $Y_k$ and $X_k$ (respectively $S_{Y}$ and $S_X$):

\begin{equation}
\label{SpectralDensity}
    S_Y(\nu)=\f{2}{\pi}\f{\sigma_Y^2}{\sigma_D^2} S_X(\nu) + 2 \frac{\sigma_Y^2}{\nu_{rep}}
\end{equation}

which shows that the digitization noise $\sigma_Y^2$ is spread as a white background over the acquisition bandwidth
$\nu_{rep}/2$.
This constant background gives the sensitivity at which $S_X$ can be measured. 
It is important to stress
 that this noise level can be squeezed down just by increasing the sampling rate.
 %which is a typical technique
% used in oversampling analog to digital converters\cite{hauser1991principles}.
As an example, we consider again the case of a sinusoidal $X_k$ such that $\sigma_D=\sigma_X$ (see 
Fig.\ref{TimeDomainFFT}c). In this case, the harmonic distortion of the noiseless $1$-bit $A/D$ converter is suppressed and replaced by a white background
in the spectrum of $Y_k$ ( Fig.\ref{TimeDomainFFT}d).

The point we want to make now is that this $1-$bit analog to digital conversion with an additive random noise is
analogous to our cavity bifurcation amplifier with thermal and quantum noises taken into account. The 
effects of thermal and quantum noises should be seen as the addition of a gaussian random variable $D_k$ to $X_k$ prior to thresholding.
The $D_k$ are assumed to be independent since the sampling interval is much larger than the reset time of the detection process. The probability distribution of the noise $D_k$ is directly related to 
the switching curve (see Fig.\ref{Fig2}f), and can be obtained from the Dykman model in both the
thermal and quantum regimes.
%As expected, the threshold is no longer "sharp" but has some finite width which
%is $\sqrt{\sigma_D^2+\sigma_X^2}$. 

%As a consequence, we can apply all the preceding framework to our CBA detector.
%Assuming that the $\nu_0$ fluctuations are generated by an underlying stationnary random process,
%most of the information (but not all) is contained in the second order statistics or equivalently the spectral density,
%the phase of these fluctuations is not relevent.

\section{Experimental results}

\begin{figure}
\includegraphics{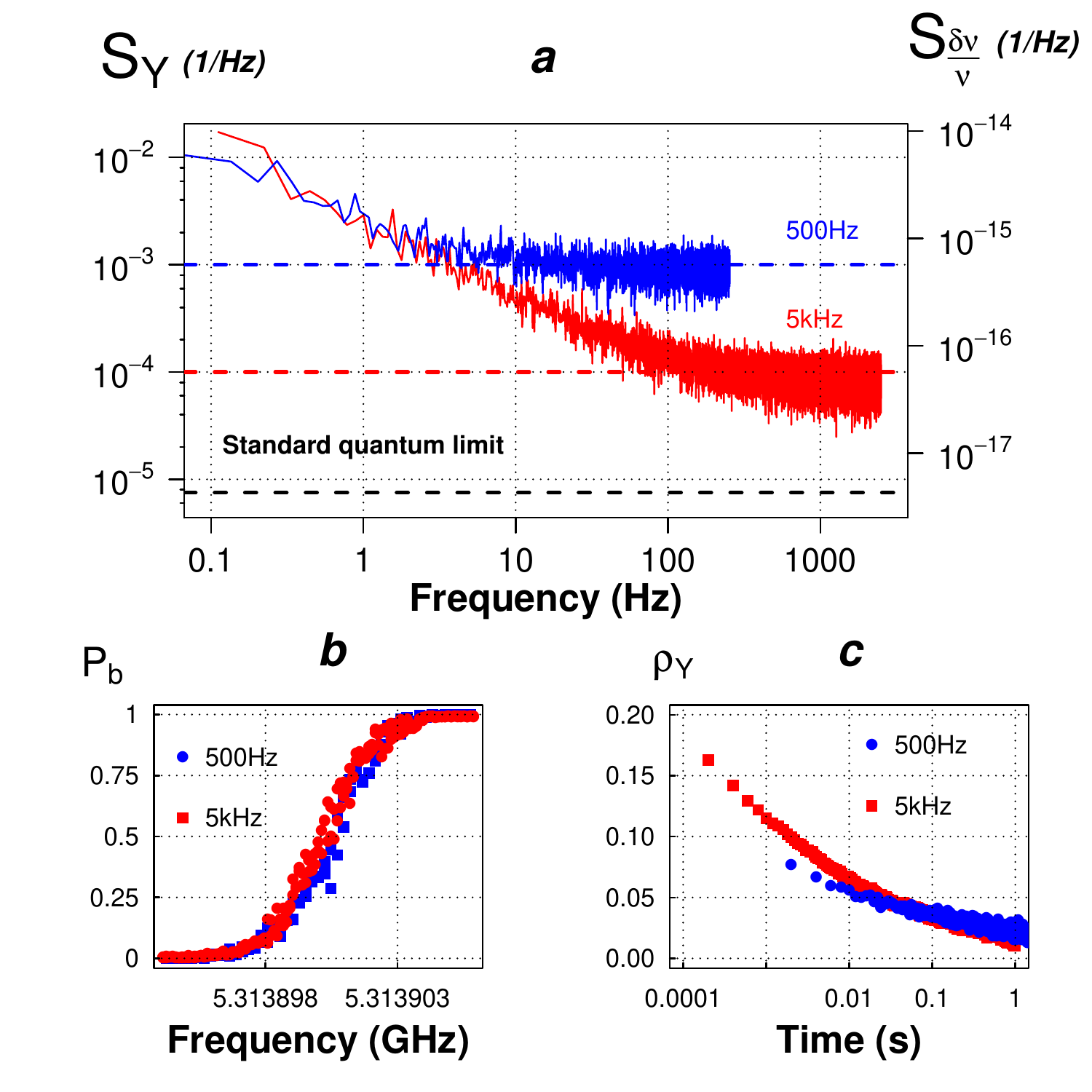}
\caption{\textbf{a)}: The spectral density of the switching signal for two repetition rates: $5$kHz (red) and $500$Hz (blue). Right scale: equivalent relative frequency jitter calculated using Eq.~(\ref{SpectralDensity}). The white background is indicated with dashed lines for both repetition rates (red: $5$kHz, blue: $500$Hz) and is consistent with the expected digitization noise.
The standard quantum limit is displayed (black dashed line). 
\textbf{b)}: 
 Switching probability curves at $5$kHz (red) and $500$Hz (blue)
repetition rates as a function of the microwave driving frequency. Each point is calculated over $1000$ events
(thus $0.2$ or $2$s of acquisition). The average $10\%-90\%$ width is $4.5$kHz or $1$ ppm of the resonance frequency of the
resonator. \textbf{c)}: autocorrelation of the switching signal. Note the log scale on the time axis.
}
\label{Fig3}
\end{figure}

We set the working point of our experiment to $p_{exp}\approx 1/2$
and then record the outcome of the CBA as a binary array over a time $\tau \approx 3$min at a
temperature of $11$mK $\pm 2$mK (measured with a PdFe magnetic susceptibility thermometer) and
for two different repetitions rates: $500$Hz and $5$kHz. 
We first note that we do not see any dependence of the switching curves on the repetition rate, which allows us to exclude heating effects as a source of correlations (see Fig.\ref{Fig3}b). 
The Spectral Density $S_{Y}$ is then computed from the array $\{Y_1,Y_2,...,Y_N\}$ (Fig.\ref{Fig3}a)
using a Fast Fourier Transform routine. 
%The dependance of the autocorrelation with microwave power was also studied and did not show any variation, however
%the available dynamical range for power variations is only one order of magnitude for the bifurcation phenomenon.
From the experimental value of $\langle \delta Y_{k+1}  \delta Y_k  \rangle \approx  0.16 \sigma_Y^2$ (extracted from Fig.\ref{Fig3}c),
we obtain the ratio $\sigma_X/\sigma_D \approx 0.5$ and from the experimental width of the switching curve
we have $\sqrt{\sigma_X^2+\sigma_D^2}=4.5$kHz. We thus deduce $\sigma_D \approx 4.0$kHz, which
allows one to convert $S_Y$ to a fractional frequency noise spectrum $S_{\delta \nu/\nu}$ (shown on the right scale of Fig.\ref{Fig3}a).  Note that the value of 
$\sigma_X \approx 2$ kHz $\approx 0.4$ ppm of the resonant frequency of the mode  is comparable to the state of the art in superconducting quantum bits achieved with 3D cavities~\cite{Paik3DCavityPRL2011}.
As expected, the white background noise corresponding to the digitization noise is present, its level agrees well with
 the prediction and can be squeezed down by increasing the repetition rate. 
%   In addition, the fluctuations of this background due. usual fluctuations of the estimated spectrum can be reduced by
%  averaging
%   The signal over noise ratio of an FFT is $1$! FFT calculation gives an estimate of the power spectral density
%   at a given frequency whose relative standard deviation is $100\%$!!!
%   To decrease this noise, adjacent frequency averaging or multiple spectra averaging are required.
%   Evaluation noise of the noise.
  This white background gives us the sensitivity of the spectrum measurement. Using the theoretical prediction
 for $\sigma_D$ in the quantum regime~\cite{Tancredi1} we can rewrite this white background noise as:
\begin{equation}\label{MinNoise}
S_{\delta \nu/\nu_0}^{theo} = \frac{3^{2/3}}{2^{5/3}} \frac{\gamma^2}{\nu_0^2} \f{ \sigma_Y^2}{ n_c^{4/3}}\frac{1}{\nu_{rep}}
\end{equation}
where $\nu_0$ is the resonant frequency of the CBA.
This background digitization noise is plotted on Fig.\ref{Fig3}a for the two repetition rates $500$Hz and $5$kHz.
It is interesting to compare this sensitivity to a fundamental scale
which is the standard quantum limit of a weak continuous measurement of the frequency
of a resonator~\cite{ClerkRMP2010} in comparable experimental conditions: average photon number
in the cavity (here $\overline{n} \approx 2 n_c\approx 500 $) leaking at rate $2\pi\gamma$ (here $\approx 2$ MHz).
The frequency fluctuations of the resonator equivalent to the shot noise of the driving
coherent state are given by $S_{\delta \nu/\nu}^{sn} =2\nu_0^2/(\gamma^2 \dot{n})$ where $\dot{n}=2\pi \gamma\overline{n}$.
Remarkably, for the maximal theoretical repetition rate of this detector ($\nu_{rep}\approx \gamma/5 \approx 50$kHz)
 the theoretical prediction for the sensitivity of the bifurcation as a noise spectrum analyser
would be comparable to the standard quantum limit. Experimentally, we used a maximum repetition rate of $5$kHz,
giving a sensitivity within an order of magnitude of the standard quantum limit.
%Note that this standard quantum limit can be beaten with tricks like the one described in ref\cite{Yan2012}
% or by calculating first the autocorrelation, setting the zero time value to the first time step value.
% There are no restriction a priori on the value the sensitivity can take for a spectrum measurement, this is of course 
%not the case for a linear amplifier

%Such a sensitivity makes the bifurcation a very interesting tool for

In addition to the digitization noise, a significant $A/f$ frequency noise is present in our sample 
with $A \approx 10^{-15}$. From flux modulation measurements\cite{Tancredi1}, we can put an upper bound on the contribution of flux noise at the optimal working point ($\phi=0$ where sensitivity to flux noise is only second order), and show that it has
 negligible contribution. In addition, because of the small value of the participation ratio 
 $L_{SQUIDs}/L_{tot} \approx 2.5 \%$ (where $L_{SQUIDs}$ is the total inductance of the SQUID array, and $L_{tot}$ the total inductance of the cavity), critical current noise has also negligible contribution. Finally, as the noise amplitude observed is compatible with previous observations made in Kinetic Inductance Detectors~\cite{gao_noise_2007,gao_experimental_2008,gao_semiempirical_2008}, we conclude that 
 dielectric noise is probably the source for the observed $1/f$ noise in this device.
% and a temperature study of the frequency of similar design and fabrication resonators show the typical signature of two levels dielectric fluctuators. 

%%%%%%%%%%%%%%%%%%%%%%%%%%%%%%%%%%%%%%%%%%%%%%%%%%%%%%%%%

\section{ Conclusion}

We have presented a model that provides a deeper insight into threshold detectors.
This model allows direct access to the spectral density of any noise source coupled to such detectors and is reminiscent of noise shaping with "dithering" in analog to digital conversion. 
It was applied to measure the
 frequency fluctuations of a Cavity Bifurcation Amplifier demonstrating the presence of a $1/f$ noise whose amplitude is compatible with previous observations of dielectric noise in Kinetic Inductance Detectors. 
The main advantage of this technique as an on-chip detector, is its dispersive nature which avoids the dissipation and backaction associated with the voltage state of a SQUID amplifier or switched hysteretic junction.
 This allows a lower thermalization temperature
of the degrees of freedom considered.
 The sensitivity of this technique as a noise spectrometer is potentially of the order of the standard quantum limit of a weak continuous measurement.
The potential of this technique for the extensive characterization of decoherence sources in superconducting quantum
bits circuits is thus high. It could provide in situ measurement of noises of any origin, including magnetic, charge, critical current, dielectric, kinetic inductance noises. They can be measured most effectively if the coupling is tunable.
In addition, the detection bandwidth of this method is half the repetition rate which is in our case limited by the reset 
time of the bifurcation detector. 
A lower quality factor than that used in our experiment could allow repetition rates of order of several hundreds of MHz. Obtaining the noise spectrum over this frequency range with a lower digitization noise would be of great interest.
Finally, we note that only partial information on a random process is provided by the second order statistics. As a consequence, it would be interesting to generalize this method to higher order correlators. Apart from qubit diagnostics, the technique may have important applications for the measurement of the full counting statistics of a quantum conductor~\cite{Blanter2000}.

We wish to thank D. Est\`{e}ve and all at the Quantronics Group at CEA Saclay for their support over many years, especially P. Bertet and A. Palacios-Laloy who fabricated the sample, A. Tzalenchuk and T. Lindstr\"{o}m (NPL) for helpful discussions and the loan of equipment, and John Taylor and Howard Moore for technical help.
G. Ithier acknowledge financial support from the Leverhulme
Trust (Early Career Fellowship SRF-40311) and P. J. Meeson acknowledge financial support from the EPSRC (grants EP/D001048/1 and EP/F041128/1) and EMRP. The EMRP is jointly funded by the EMRP participating countries within EURAMET and the European Union.

%\appendix
%Noise Budget

\begin{table*}[ht]
\caption{Notations}
\centering
\begin{tabular}{l l}
\hline \hline
Symbol & Definition / Result  \\ [0.5ex]  % Table heading
\hline
$D_k $ 		& Added noise prior to thresholding at time step $t_k$ considered as a random discrete
	 		variable. \\
$n_{ph}$  &  Photon number in the third harmonic mode of the superconducting cavity. \\
 $\nu_{rep}$    & Repetition rate of the acquisition process (typically up to a few kHz).  \\
 $P_X(x), P_D(d)$ 	& Probability density of the random variables $X$ and $D$.\\
$P_{X,X}(x_1,x_2;t)$ & Joint probability density of the random stationary  process $X_t$. \\
$\mathbb{P}(\omega_1/\omega_2)$ & Conditional probability for event $\omega_1$ to happened
			 knowing that event $\omega_2$ has happened. \\
$p(x)$ 		& Shorter notation for the conditional probability : $\mathbb{P}(Y_k=1/X_k=x)$. \\
$p_{exp}$ 	& Experimental bifurcation probability, obtained by counting bifurcation
			 events over $\approx 10^3$ sampling pulses. \\
$Q$   		& Quantizer function: $Q[x]=1$ if $x>0$ and $Q[x]=0$ if $x<0$. \\
$\rho_X(t)$   & Normalized autocorrelation of the $X_t$ process: $\lan \delta X_t \delta X_0 \ran/\sigma_X^2$ \\
$\rho_Y(t_q)$ & Normalized autocorrelation of $Y$: $\lan \delta Y_{k+q} \delta Y_k\ran /\sigma_Y^2$ \\
$\sigma_X,\sigma_Y,\sigma_D$    & Standard deviations of the random variables $X, Y, D$.\\
$S_Y(\nu)$   	&  Spectral density of the binary array $\{Y_1,Y_2,...,Y_N \}$. \\
$S_X(\nu)$  	& Spectral density of the random process $X_t$.  \\
$t_k=k/\nu_{rep}$ 		& $k^{th}$ sampling time.  \\
$\lan X \ran$	& Average of the random variable $X$. \\
$\delta X=X-\lan X \ran$         &  Fluctuation of $X$. \\
$\Delta X$      &  $10\%-90\%$ width of the S-like curve: $p_{exp}$ as a function of $\lan X \ran$.\\
$X_t$      		& Input of the detector considered as a time dependent random process. \\
$X$			& Shorter notation for $X_t$ when the time dependence can be omitted. \\
$X_k$     		& Shorter notation for $X_{t_k}$, the input sampled at time step $t_k$ considered 
			 as a discrete random variable. \\
$\lan \delta X_{k+q}.\delta X_k \ran$ & Autocorrelation of the input signal of the detector. \\
$Y_k$    		& Output of the detector at time step $t_k$ considered as a discrete random variable.  \\
$\lan \delta Y_{k+q}.\delta Y_k \ran$ & Autocorrelation of the output signal of the detector. \\
\hline
\end{tabular}
\label{TableNotations}
\end{table*}

\bibliography{NoiseSpectralAnalysisCBAbiblio}

\end{document}